\documentclass[floatfix]{revtex4}

\usepackage{amssymb,graphicx}
\usepackage{epsfig}
\usepackage{booktabs}
\usepackage[usenames]{color}
\usepackage{multirow} 
\usepackage{rotating}

\begin{document}

\title{Improving the scalability of parallel N-body applications with an event driven constraint based execution model}

\author{Chirag Dekate$^{1,2}$, Matthew Anderson$^{2}$, Maciej Brodowicz$^{2}$, Hartmut Kaiser$^{1,2}$,Bryce Adelstein-Lelbach$^{2}$,Thomas Sterling$^{1,2}$}

\affiliation{
${}^1$Department of Computer Science, Louisiana State University, Baton Rouge, LA 70803,\\
${}^2$Center for Computation and Technology, Louisiana State University, Baton Rouge, LA 70803
}

\date{\today}

%
%
\begin{abstract}
The scalability and efficiency of graph applications are significantly constrained
by conventional systems and their supporting programming models. Technology trends
like multicore, manycore, and heterogeneous system architectures are introducing
further challenges and possibilities for emerging application domains such as 
graph applications. This paper
explores the space of effective parallel execution of ephemeral graphs that are dynamically
generated using the Barnes-Hut algorithm to exemplify dynamic workloads. The workloads
are expressed using the semantics of an Exascale computing execution model called ParalleX. 
For comparison, results using conventional execution model semantics are also presented.
We find improved load balancing during runtime and automatic parallelism discovery
improving efficiency using the advanced semantics for Exascale computing. 
\end{abstract}

\maketitle

%
%
\section{Introduction}
A large class of problems in physics and molecular biology can be represented using a
particle interaction method commonly known as N-Body and computational techniques 
based on these discretization methods. The science domains utilizing the particle 
interaction discretization model are limited by the number of particles that can be 
simulated and the time it takes to execute the computational techniques. Conventional 
practices have significantly advanced particle interaction based methodologies. However, 
the combined ecosystem of emerging multicore based system architectures and conventional 
programming models are imposing grave challenges to the continued effectiveness of 
these methods. This research identifies and addresses these challenges through 
the hypothesis that emerging system architectures and extreme scale oriented 
runtime systems can dramatically improve the end science.

Applications based on graphs and tree data structures rely on more dynamic, 
adaptive, and irregular computations. This work explores an exemplar dynamic tree 
based application embodied by an N-Body simulation. Systems comprising many particles 
(N-Body problem) interacting through long-range forces have considerable computational
science interest. N-Body systems comprising three or more particles do not have a closed 
form solution; consequently, iterative methods are used to approximate solutions 
for the N-Body problem. The N-Body problem simulates the evolution of $n$ particles 
under the influence of mutual pairwise interactions through forces such as 
gravitational pull or electrostatic forces. This work focuses on gravitational 
forces operating on the N-Body system and the Barnes-Hut approximation of the N-Body solution.

While several approaches to simulating N-Body systems exist, the Barnes-Hut 
algorithm~\cite{barneshut} is widely used in astrophysical simulations 
mainly due to its logarithmic computational complexity while generating results that are 
within acceptable bounds of accuracy. In the Barnes-Hut algorithm the particles
are grouped by a hierarchy of cube structures using a recursive algorithm which subdivides the
cubes until there is one particle per sub-cube. It then uses the adaptive octree
data structure to compute center of mass and force on each of the cubes. The data
structure organization is key to the resultant $O(N\log{N})$ computational complexity
of the algorithm.

During the simulation the position of the particles change as a result
of interaction with other particles, consequently the representative Barnes-Hut tree 
changes dynamically during the computation. The dynamic tree based evolution of the 
system is not as common to conventional computational applications, which are
dominated by simulations where the 
fundamental data structures are arrays/matrices that can exploit data locality for performance, 
through techniques such as row/column striping~\cite{striping}.
Graph/Tree data structures have non uniform data access patterns
which prevent tree based applications from taking advantage of data locality and 
fully utilizing the parallelism offered by the underlying system. 
The emergence of multicore processor architecture, while providing capabilities
for significant intra-node parallelism, poses new challenges.
When used in isolation, traditional programming techniques 
fail to expose the underlying parallelism to the high performance 
computing applications since they cannot address the intra-node parallelism 
provided by multicore processor architecture. Conventional clusters comprise
multicore nodes interconnected by a low latency high bandwidth network such as 
InfiniBand and are usually
programmed using MPI~\cite{MPISpec}. When conventional techniques such as message passing 
are used in isolation the intra-node parallelism offered by each of the multicore nodes 
is under utilized, primarily due to limitations of the implementations of MPI. Performance 
and scalability of graph applications are further affected by high performance computing 
system architectures that are more suitable for computations with static, predictable data access 
patterns. Systems architectures such as clusters and massively parallel processors (MPPs) have co-evolved
with programming models and applications that provide high data locality to computations.

The fundamental challenges to improving performance and scalability of parallel N-Body 
simulations using the Barnes-Hut algorithm in a system context are:

\begin{itemize}
 \item \textbf{Dynamic Load Balancing} The Barnes-Hut algorithm generates a new 
unique tree for each iteration which is different
than the tree used in the previous iteration; consequently,
conventional load balancing techniques such as striping are ineffective and 
result in poor load balancing. 
 \item \textbf{Variable Workload} In addition to the varying Barnes-Hut tree, 
a fundamental resource management problem
encountered while simulating these systems is the variable workload 
per particle. That is, each particle
in the system will interact with a constantly varying number of particles in the system. Consequently, predicting
workload per particle in N-Body simulations is difficult.
 \item \textbf{Data Driven Computation} The Barnes-Hut tree algorithm, like 
other graph computations, is data driven. The computations
performed by the Barnes-Hut algorithm are dictated by the vertex and edge 
structure of the tree rather than by directly
expressing computations in code. This results in irregular communication patterns and
irregular data access patterns limiting ability to parallelize using conventional techniques.
 \item \textbf{Data Locality} The irregular and unstructured computations in dynamic graphs result in poor data
locality resulting in degraded performance on conventional systems which rely on
exploitation of data locality for their performance.
 \end{itemize}

This paper compares the Barnes-Hut tree algorithm using the conventional execution model 
semantics of OpenMP~\cite{openmp1,openmp2} with the dynamic data-driven semantics and
execution techniques of the ParalleX~\cite{scaling_impaired_apps,tabbal} execution model.
In section~\ref{sec_parallex}, an overview of the ParalleX execution model will be given
along with the key characteristics that distinguish it from 
the communicating sequential processes execution
model typified by MPI~\cite{MPISpec}. In section~\ref{barneshut}, the Barnes-Hut algorithm will be 
recast into dynamic data-driven semantics for use in an Exascale execution model. 
Section~\ref{experiment_setup} will
present the experimental set-up and tests performed while section~\ref{results} will present
the results of the experiments. Our conclusions are found in section~\ref{conclusion}.

%
%

\section{ParalleX}
\label{sec_parallex}
As technologies advance towards enabling Exascale computing, major challenges
of scalability, efficiency, power and programmability will need to be addressed.
New models of computation will be needed to provide the governing principles
for co-design and operation of the cooperating system layers from programming
models through system software to process or architectures.

The ParalleX execution model~\cite{scaling_impaired_apps,tabbal} 
is derived with the goal of specifying
the next execution paradigm essential to exploitation of future
technology advances and architectures in the near term as well as
to guide co-design of architecture and programming models in
conjunction with supporting system software in the long term.
ParalleX is intended to catalyze innovation in system structure, operation,
and application to realize practical Exascale processing capability by
the end of this decade and beyond.

The four principal properties exhibited by ParalleX to this end are:
\begin{enumerate}
\item Exposure of intrinsic parallelism, especially from meta-data, 
to meet the concurrency needs of scalability by systems in the next decade,
\item Intrinsic system-wide latency hiding for superior time and power efficiency,
\item Dynamic adaptive resource management for greater efficiency by exploiting runtime information, and
\item Global name space to reduce the semantic gap between application requirements and system functionality both to enhance programmability and to improve overall system utilization and efficiency.
\end{enumerate}

ParalleX provides an experimental conceptual framework to explore these goals and
their realization through the concepts identified above to inform the
community development of a future shared execution model. The evolution
of ParalleX has been, in part, motivated and driven by the needs of the
emerging class of applications based on dynamic directed graphs. Such
applications include scientific algorithms like adaptive mesh refinement,
particle mesh methods, multi-scale finite element models and informatics related applications
such as graph explorations. They also
include knowledge-oriented applications for future web search engines,
declarative user interfaces, and machine intelligence. The strategy of
ParalleX is to replace the conventional static Communicating Sequential
Processes model~\cite{csp} of computation with one based on a dynamic
multiple-threaded work-queue processing model discussed above. ParalleX
also incorporates message-driven computation through Parcels, an advanced
form of active messages.

ParalleX can be viewed as a combination
of ParalleX Elements and ParalleX Semantic Mechanisms. ParalleX Elements
are the fundamental building blocks of the execution model comprising
ParalleX Threads, ParalleX Parcels, Localities and ParalleX Processes.
ParalleX mechanisms are complex interactions enabled by the structure,
semantics of the ParalleX elements, and the dynamics involved in the
execution model (see Table~\ref{pxconcepts}).

\begin{center}
\begin{table*}[htbp]
   \centering
      \caption{ParalleX Concepts Summary}
   \label{pxconcepts}
\begin{tabular}{|c|c||c|}
\hline \textbf{ParalleX Elements} & \textbf{ParalleX Mechanisms} & \textbf{Emergent Properties} \\
\hline ParalleX Threads & Multi-Threaded Work Queue & Percolation \\
\cline{1-2} Parcels & Message Driven & Security \\
\cline{1-2} Local Control Objects & Synchronization & Adaptive Scheduling\\
\cline{1-2} ParalleX Processes & Active Global Address Space & Self Aware \\
\cline{1-2} Localities & Overlap Computation and Communication & Split-Phase Transactions \\
\hline
\end{tabular}
\end{table*}
\end{center}

ParalleX Elements are the fundamental building blocks managed by the
mechanisms. The elements comprise: ParalleX Threads, Parcels, Processes,
LCOs, and Localities. Each of the elements will be discussed below. 
More discussion of ParalleX mechanisms and emergent properties, including message-driven 
computation, the Active Global Address Space, percolation, 
and split-phase transactions, is found in~\cite{chiragsthesis}.

\subsection{ParalleX Threads}
\label{sec_pxthreads}
The fundamental element that accomplishes computational work is, as is referred to in many programming models,
the ``thread''. Like more conventional threads in other models, ParalleX
confines the execution of a thread to a single localized physical domain,
conventionally referred to as a ``node'' and here as ``locality'' in ParalleX
terminology (see section~\ref{subsec:localities}). A thread is a place of local control and data for execution
of a set of instructions.

Unlike traditional threading models such as Pthreads~\cite{pthreads}, 
in ParalleX threads are highly ephemeral. 
A ParalleX thread may contain a continuation~\cite{continuations},
which specifies an action to be taken once the current thread terminates.

ParalleX threading is also heterogeneous. ParalleX threads of many
different forms (code) can be instantiated at the same time without
restrictions. Both these properties distinguish ParalleX threads from
those of UPC~\cite{upc1,upc2}. 

A particularly important distinguishing aspect of ParalleX threads is that they are first class
objects; they are named as is any other variable or object and may be
manipulated within the limitations of their type class from any
part of the allocated system. ParalleX threads
may be as short as a atomic memory operation on a given data structure
element or as long as a conventional process that lasts the entire
duration of the user program. The intended mode of operation, however, is
for short lived threads that operate exclusively on private or local
data within a single locality and then pass on the flow of control to a remote
locality (by instantiating a new thread there), where the execution continues based on
data placement. 
 
A ParalleX thread which is unable to proceed in its execution due to incomplete control state
transitions (e.g., a required intermediate value is still pending) is
``suspended'' (it has no more work it can do) and is transformed in to a new kind of 
object with the same name. This is one of a general class of lightweight objects that includes
control state referred to in ParalleX vocabulary as a ``Local Control Object''
or ``LCO'' described in section~\ref{subsec:lcos}. In doing so the ParalleX
thread saves its requisite private state and relinquishes the physical
resources for other active threads in the locality. A ParalleX thread differs
in one last way from more conventional threads in that typically a thread
is assumed to be sequential in its instruction issue. ParalleX incorporates
a possible intra-thread parallelism based on a data flow control schema
for intermediate values not unlike static dataflow~\cite{dennisdataflow}. This eliminates
anti-dependencies at least for intermediate (intra-thread) values
allowing generalized application to a diverse set of processor core
instruction set architectures, thus facilitating the use of heterogeneous
computing systems. Overall, ParalleX threads support parallelism at two levels:
concurrent threads and operation level parallelism within a thread through
the data flow control.

\subsection{ParalleX Parcels}
\label{sec_parcels}
ParalleX embodies a powerful, general, and dynamic method of
message-driven computation called Parcels that allows the invocation of threads on
remote localities through a class of messages that carry information
about an action to be performed as well as some data that may be
involved in the remote computation.
Parcels (PARallel Control ELements) are a form of active
message~\cite{activemessages} that extend
the semantics of parallel execution to the (highly asynchronous) inter-locality domain for
scalability and efficiency. The parcel supports message-driven
computation allowing work to be moved to the data rather than the data
always gathered to the localized site of the work. Parcels reduce both
bandwidth requirement and hide latency for greater efficiency and
scalability. Parcels extend the effectiveness of the work-queue model
to operate across localities, not just within a single locality, as
would be the case without them.

\subsection{Processes}
Conventionally, a process is a form of a task or action comprising many
specified steps or instructions. A process generally runs on a processor
core and perhaps communicates with other processes via messages in the
system I/O distributed name space. ParalleX views processes as contexts
containing named entities including data constants and variables,
concurrent threads both active and suspended, local control objects, and
other child processes. Processes also enable ``symmetry'' of semantics
that allow a user to perform remotely the same activities that an application
would perform locally. The important property of a ParalleX process
that distinguishes it from conventional process oriented models is that it can
occupy many localities (i.e. system nodes) at the same time
using the combined resources of the collection of assigned localities to
perform the process' active threads and to store the process data. Like
threads, processes are ephemeral. A process is instantiated from a combination
of a procedure specification, a set of operand values,
and allocated physical localities at any point in time of the execution
of a parent process. 

The ParalleX semantic mechanism associated with processes is the Active Global Address Space (AGAS).
ParalleX supports a shared memory model that distinguishes it from the
conventional distributed memory MPP and cluster system architectures,
but is also different from the cache-coherent shared memory systems such
as symmetric multiprocessors (SMPs). It incorporates a strategy among the class of ``Global Address
Space'' models sometimes referred to as put-get models. AGAS differs
from the more widely employed model called Partitioned Global Address Space (PGAS~\cite{gasnet,UPCSpec,Charles:2005:XOA:1103845.1094852})
in one specific and important way. With PGAS a virtual object has a restricted scope of
allocation within the confines of a particular partition.
ParalleX, at the risk of additional overhead, relaxes this constraint. It
permits virtual objects to be moved across the system between localities
while retaining their original virtual addresses. This is the ``active'' aspect
of AGAS versus the passive or static nature of the ``partitioned''
aspect of PGAS. This property is critical for dynamic directed graphs and for
load balancing as well as system reconfiguration.

\subsection{Local Control Objects}
\label{subsec:lcos}
One of the most powerful aspects of the ParalleX model is the way it
represents and manages global flow control. Typically, distributed
applications employ the construct of the global barrier to manage phased
execution in what is referred to as the Bulk Synchronous Protocol or
BSP model~\cite{bsp}. Simply, BSP uses a compute-synchronize-communicate
phased management approach where the exchange of data via messages only
occurs after all processes have finished their respective work for the
current cycle. A global barrier is a synchronization object that ensures
the strictness of this ordering requiring all processes to check in to the
barrier upon completion of their work and to not proceed with their
communication phase until notified to do so by the barrier
(which makes certain that all processes have finished). For many
applications and systems this has proven an effective approach. But
this coarse grained parallelism precludes the more fine grained and
adaptive parallelism required for extended scalability.

Local Control Objects (LCOs) are event driven conditional structures
that, upon satisfying specific criteria, will cause a thread or some other
action to be
instantiated. Events are either direct accesses by threads within the
same locality and process, or through parcel messages outside these
restricted domains. Local Control Objects improve efficiency by avoiding
barriers and permitting highly dynamic flow control.

Among the simplest form of an LCO is a mutex~\cite{mutex}.
But the more interesting forms include the dataflow template and the
futures~\cite{futures2} construct. A future
is a promise of a returned value, which may be any first class object.
If the value itself is not required, such as manipulating the metadata
surrounding it, then computation may proceed. Further computation will be
blocked only when the actual value is required and not just its placeholder. 
But futures are far more powerful. They can coordinate
anonymous producer-consumer computation such as histogramming and
can also coordinate access to a shared structure such as a vertex
of a graph.

\subsection{Localities}
\label{subsec:localities}
Like many models dealing with distributed concurrent computing, ParalleX recognizes a
contiguous local physical domain. In the case of ParalleX, a locality is the locus of
resources that can be guaranteed to operate synchronously and for which
hardware can guarantee compound atomic operation on local data elements.
Within a locality, all functionality is bounded in space and time
permitting scheduling strategies to be applied and structures
that prevent race conditions to be built. The locality
also manages intra-locality latencies and exposes diverse temporal locality attributes.
A locality in ParalleX is a first-class object, meaning that each locality in a ParalleX
system has a globally unique identifier associated with it and can be addressed throughout
the user global address space. The locality comprises heterogeneous processing
units including, for example, a combination of homogeneous multicore processing elements and GPGPU
accelerators.

\begin{center}
\begin{table*}[htbp]
   \centering
      \caption{Overview of ParalleX mechanisms that enable dynamic directed graph processing. }
   \label{tabpxdg}
   \begin{tabular}{|l|l|} 
\hline
      \toprule
      \textbf{Dynamic Directed Graph Requirements }  & \textbf{ParalleX Enablers} \\
      \midrule
\hline
      Data-driven computing & Parcels carry data, target address, action, continuation.\\
                            & Allows work to be moved to data.\\
                            & Enables migration of flow control.\\
                            & Supports asynchronous event driven execution.\\
\hline
      Support for fine-grained parallelism & Work queue model enables split phase transactions, \\
                                           & i.e., overlapping computation with communication. \\
                                           & Threads support dataflow on transient values. \\
\hline
      Dynamic thread support & Multithreaded light weight entities; \\
                             & first class objects;  \\
                             & may be created and addressed in suspended state; \\
                             & may be compiled to different types of processors. \\
\hline
     Global memory access & PX communication using AGAS and Parcels; \\
                          & non cache-coherent global memory space; \\
                          & supports copy semantics and affinity relationships.\\
\hline
      \bottomrule
   \end{tabular}

\end{table*}
\end{center}

The ParalleX mechanisms that enable the dynamic directed graph processing needed for
the Barnes-Hut algorithm are summarized in Table~\ref{tabpxdg}. The next section describes
in detail the Barnes-Hut algorithm as implemented in ParalleX.

%
%

\section{Barnes-Hut Algorithm in ParalleX}
\label{barneshut}
\begin{figure}[htp]
\centering
\includegraphics[scale=1.0]{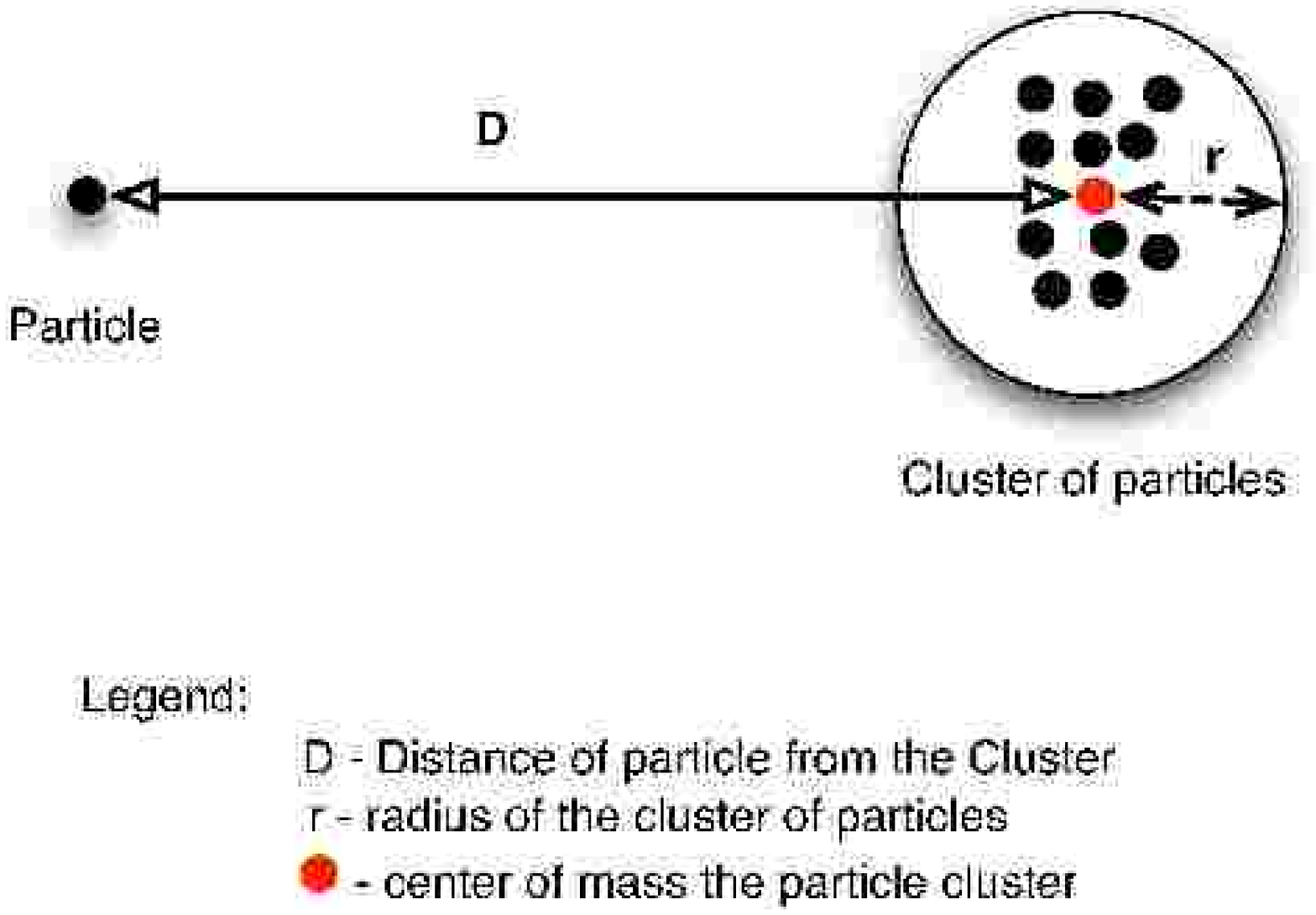}
\caption{Barnes-Hut Approximation Criteria}
\label{thetafig}
\end{figure}

The Barnes-Hut algorithm~\cite{barneshut} is widely used in
the astrophysics community as it is
the simplest hierarchical N-Body algorithm. The N-Body algorithm
exploits the same idea used by all tree codes where the forces on a
body from a remote cluster of bodies can be approximated by
treating the remote cluster as a single body. The accuracy of these
approximations (see Figure~\ref{thetafig}) depends on the distance (D) of the cluster from the body
and the radius (r) of the cluster of particles. Remote clusters of bodies
are treated as a single body only if D is greater than $r/\theta$, consequently
the parameter $\theta$ controls the error of the approximation.

The Barnes-Hut algorithm has been successfully parallelized using several 
techniques~\cite{warren, salmon_thesis, liu, psingh, dubinski, makino}. The key challenges in parallelizing
the Barnes-Hut algorithm include domain decomposition of the Barnes-Hut tree across the
allocated memory resources and load balancing the workloads across the allocated processors.

The ParalleX version of the Barnes-Hut algorithm explored here is implemented using 
the HPX runtime system~\cite{scaling_impaired_apps}, the reference C++ implementation of
ParalleX. The ParalleX Barnes-Hut implementation has the following goals:
\begin{itemize}
\item Parallelizing force calculation step using HPX
\item Creating interaction list and co-locating related work to provide data locality
\item Utilizing futures based asynchrony and work queue based computation for force calculations
to provide balanced loads to the processors.
\item Providing support for controlling the granularity of each HPX Thread to determine the best
workload characteristics for fast, scalable processing of the algorithm.
\end{itemize}

Inherent in the Barnes-Hut algorithm is a global barrier at the tree construction stage 
of every iteration, where
the algorithm requires positional information of all particles in the system. The research implementation
explores opportunities to exploit parallelism within each iteration. The computationally intensive force
calculation is the key focus for parallelization using the HPX runtime system. The main challenges in
parallelizing the force computation are:
\begin{itemize}
 \item exposing fine grained control and parallelism at the particle interaction level and devising methodology
to express each interaction as a work unit
 \item the number of interactions for each particle varies across iterations; consequently, new ways of
providing dynamic load balancing are required to manage efficient execution of the fine grained computations.
\end{itemize}

\begin{figure}[htp]
\centering
\includegraphics[width=0.8\linewidth]{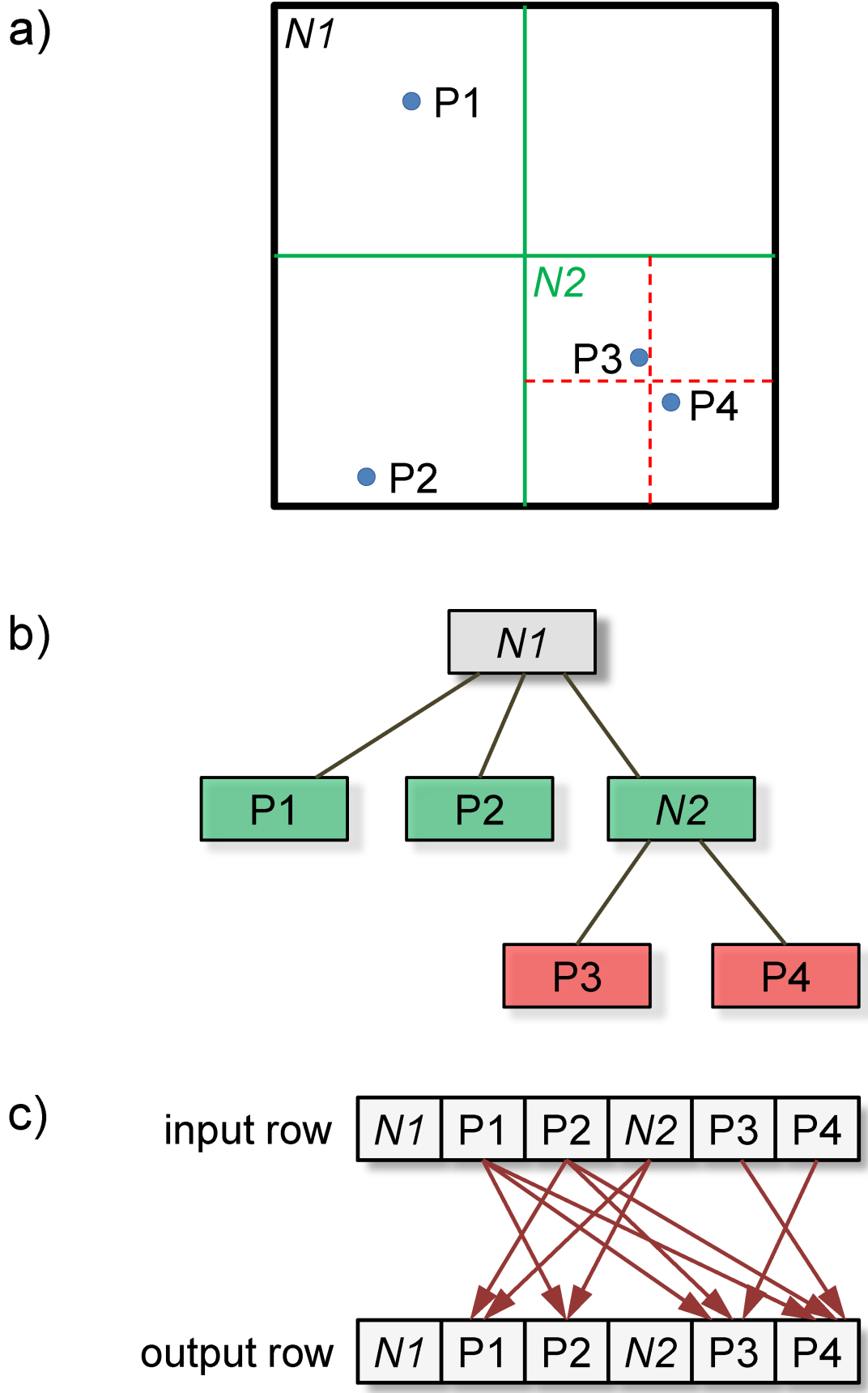}
\caption{Representing Barnes-Hut algorithm in HPX:
(a) illustrates a sample 2-D problem domain with four particles, recursively
partitioned into quadrants containing at most one particle;
(b) depicts the resultant Barnes-Hut quadtree;
(c) shows the flattened tree data converted into a row of dataflow elements 
along with data dependencies necessary to compute the next iteration values
stored into the output row.
Note that due to clustering, computation for particles P1 and P2 requires 
only two inputs as P3 and P4 may be replaced by an equivalent centroid with
parameters stored in an element corresponding to the tree node N2.}
\label{bh2hpx}
\end{figure}

The technical approach in addressing the above challenges involves exploiting a dataflow based framework for
executing the Barnes-Hut N-Body simulation. 
The dataflow approach allows for representation of each of the interactions
as a work unit and helps take advantage of a proven approach to providing scalability for scaling constrained
applications. Inherent in the HPX based dataflow approach is the use of futures for resolving data dependencies.
This provides asynchronous access to data required for computing forces
acting on each particle. Two modes of granularity
support are enabled: workload based HPX thread allocation, which allows for managing the size of HPX threads and
increasing the number of threads based on the workload; or, alternatively, workload based on a fixed number of 
threads and the variable workload size adapted to match the resource
allocations. These methodologies provide a path for implementing dynamic workload management and load balancing.

\begin{figure}[htp]
\centering
\includegraphics[width=1.\linewidth]{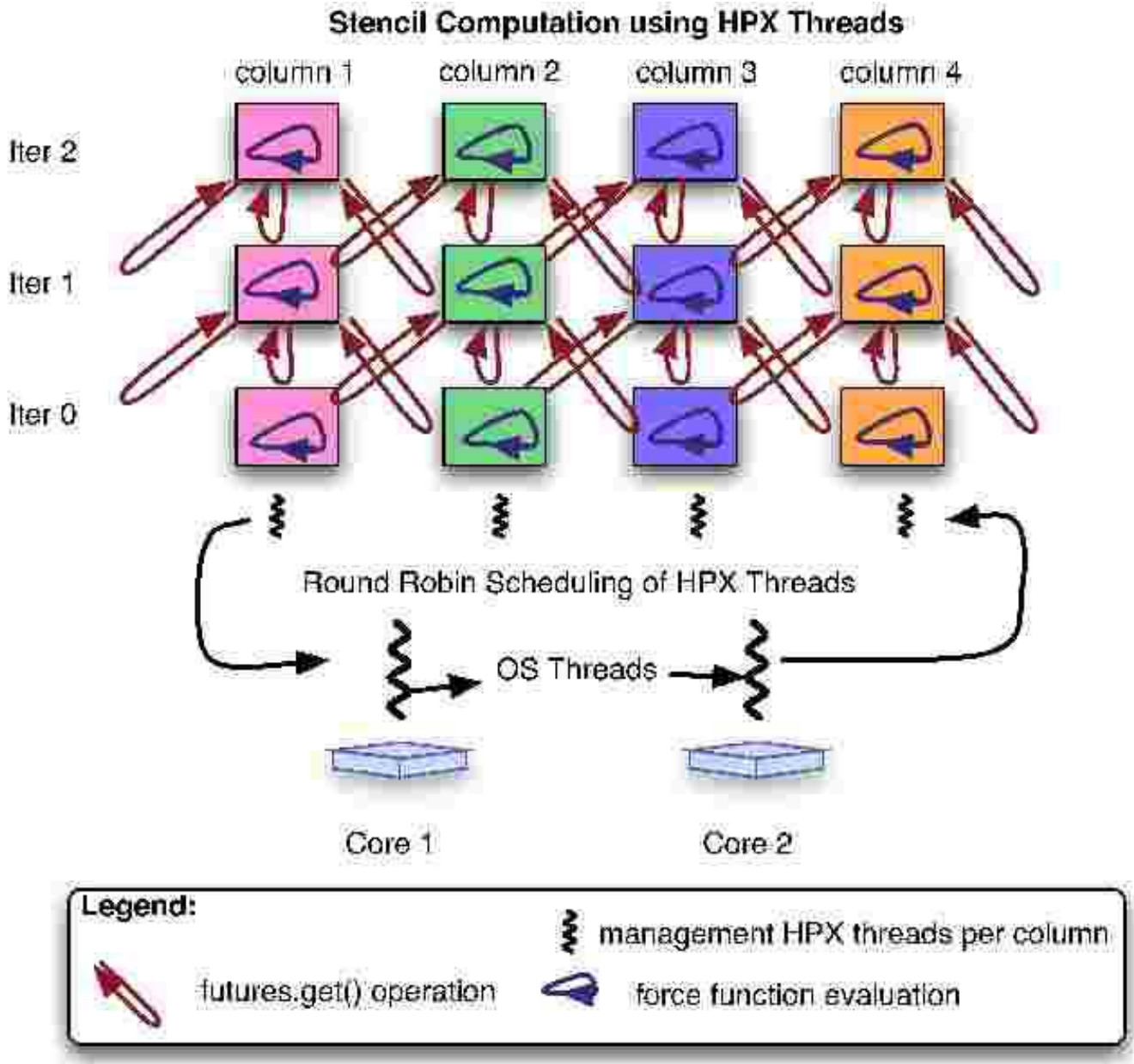}
\caption{HPX runtime system management of the Barnes-Hut algorithm workload.
The fully populated Barnes-Hut tree is flattened into an input row of 
dataflow elements, comprising all the tree nodes including the particles and the
intermediate tree nodes. A complementary output row of elements representing the result buffer is created to store
the results from the force calculation stage. Each element in both the input and output rows has a Global IDentifier (GID) associated
with it. Each dataflow element in the input row maps to a corresponding
element in the output row. Additionally, for each element in the output 
row that is of the particle type, connections are made to the input rows.
The connections are made on the basis of the Barnes-Hut approximation criteria of $D/r < \theta$.
For simplicity, each dataflow element in the figure is shown with links to 
three elements in the relevant input row that store results from the previous 
iteration; in practice this number depends on problem geometry.}
\label{fig:hpxbhexec}
\end{figure}

The most computationally intensive part of the Barnes-Hut algorithm is the force computation stage. 
The HPX parallelization
efforts and the research implementation focuses on parallelizing the force 
computation stage of the Barnes-Hut algorithm.
This is done by representing the computations as a work flow based on 
repeatable arrangement of rows of dataflow elements. The fully populated
Barnes-Hut tree is flattened into an input row of dataflow elements, 
comprising all the tree nodes including the particles and the
intermediate tree nodes (see Figure~\ref{bh2hpx}). 
A complementary output row of dataflow elements representing the result buffer is 
created to store the results from the force calculation stage. 
Each dataflow element in both the input and output rows 
has a unique Global IDentifier (GID) associated
with it. Each element in the input row maps to a corresponding element in 
the output row. 
Additionally, for each element in the output row
that is of particle type, connections are made to the input rows.
The connections are made on the basis of the Barnes-Hut approximation criteria of $D/r < \theta$. In this context, making a
connection implies that each output row element holds a vector of GIDs 
which identifies the input row elements which provide
the necessary data for computing the force on the particle in the output row.

Each column in this connected network of rows has an HPX thread for managing the internal
force computations. This network is depicted in the Figure~\ref{fig:hpxbhexec}. When executed in parallel
a global pool of these management threads are created and made available to the allocated resources. 
HPX user level threads are allocated from the global pool in a round-robin
manner and scheduled on top of the operating system threads running on
each of the cores in the parallel system.
The user may also specify the mapping of the HPX threads to preferred local queues in the HPX system.

\begin{figure}[htp]
\centering
\includegraphics[scale=0.75]{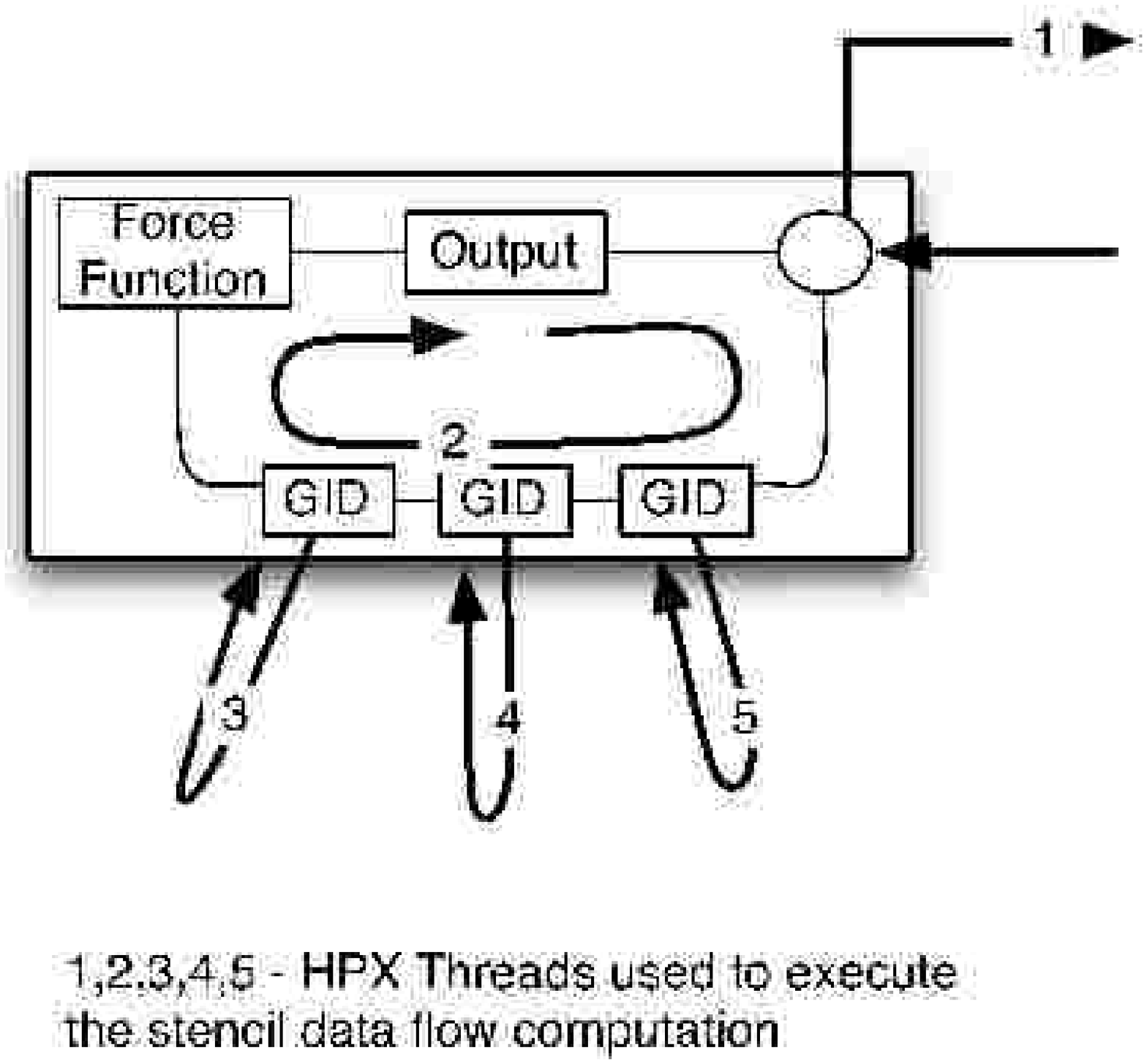}
\caption{Dataflow computation using HPX. Each HPX thread managing a column triggers the dataflow operation for
every row element allocated. The management thread is
depicted as the arrow numbered 1 in the diagram. When the
computation flow in each element is triggered, a new
HPX thread for that dataflow element is created. This thread is identified
as Thread 2 in the diagram. This element-specific HPX thread
performs a {\tt futures.get()} operation
on each of the GIDs for the input elements that the particle
interacts with. For each {\tt futures.get()} operation a new
HPX thread is created to manage the asynchronous data accesses (3,4,5).
For each input datum that is available, the local
management thread (2) computes the force function and stores the intermediate
result in a result buffer.}
\label{fig:hpxbhexec2}
\end{figure}

Each HPX thread managing a column triggers the dataflow for
every element allocated. This work flow is depicted
in the Figure~\ref{fig:hpxbhexec2}. The management thread is
depicted as the arrow numbered 1 in the diagram. When the
computation flow in each dataflow element is triggered, a new
HPX thread for that element is created. This thread is identified
as Thread 2 in the diagram. This element-specific HPX thread
performs a {\tt futures.get()} operation
on each of the GIDs for the input elements that the particle
interacts with. For each {\tt futures.get()} operation a new
HPX thread is created to manage the asynchronous data accesses (3,4,5).
For each input datum that is available, the local
management thread (2) computes the force function and stores the intermediate
result in a result buffer. For each element evaluation a total of $(2 + n)$ threads
are required (1 column manager thread + 1 local computation thread + n {\tt futures.get()}
created threads). Once the current element is completely evaluated, the next 
one is loaded and the same workflow is repeated.
The average overhead for creation of a lightweight thread in HPX is about 3-5 $\mu$s.

The futures based asynchrony in representing the interconnections provides rich
semantics for exploiting the parallelism available within each iteration.
HPX semantics and the implementation
also allow specifying granularity, or quantum of work, of each execution block, 
thus allowing for co-locating related computations
to provide implicit data locality and amortize thread overhead.

The key algorithmic difference between the ParalleX version of Barnes-Hut and 
the OpenMP version is that the ParalleX version includes the creation 
of a tagged tree data structure which is flattened to produce an interaction list. This interaction list
is then used to set up the dataflow LCO used for the force computation.

%
%

\section{Experimental set-up}
\label{experiment_setup}

All experiments have been executed on a
32 core shared memory multiprocessor system. The system has 8 processor 
sockets, each attached to a local memory bank. The processors
use HyperTransport interconnect to access memory
across the system. Sockets are populated with quad-core 2.7 GHz AMD
Opteron processors (family 8384) with 512 KB L2 cache per core and
6 MB of shared L3 cache. 
Each socket in the system is also equipped with 8 GB dual channel DDR2 DRAM. 
The system has therefore an aggregate memory of 64 GB that can be addressed 
by each of the 32 cores in the system, albeit not with a uniform latency (NUMA).
The OS used is 64 bit Linux kernel version 2.6.18.\\

All programs used in the experiment have been
compiled using the standard GNU g++ compilers version 4.4.2. The HPX
runtime system is currently under active development. For the purposes
of this experimentation, HPX version 3 is used. The OpenMP linking is
done through the version of OpenMP primitives available in the GNU g++
compilers version 4.4.2.
The OpenMP comparison code is characterized in~\cite{chiragsthesis}. The
OpenMP, HPX, and serial version of the N-Body codes are available upon 
request~\cite{nbody_download}.

The input data used in the problem size variation studies have been generated using
the standard Plummer model~\cite{plummer} which is used to model globular clusters.
The program used to generate the input data was obtained from the 
online computational course created by Piet Hut~\cite{directnb}.

In the experiments conducted here, the problem sizes are varied between 10,000
and 100,000 particles. For each of the problem sizes, the number of operating system threads 
are varied and the grain sizes are varied to determine the correlation of the parallelism to 
the workload granularity. For each of the combinations the time taken to execute the force calculation 
loop is measured. 

%
%
\section{Results}
\label{results}

In this section, results from the HPX dataflow approach to Barnes-Hut and an OpenMP Barnes-Hut implementation 
are presented. Parallelism in the HPX Barnes-Hut code
was controlled using two main parameters: the number of operating systems threads and the
workload grain size (which directly impacts the number of HPX threads used). As the grain size
of the workloads decreases (the work per thread is made more fine-grained) the number of HPX threads 
increase and the work done by each of the HPX threads decreases. The number of operating system threads 
controls the explicit parallelism
such that each of the threads gets the activities to be performed from the pool of HPX threads created.
The maximum number of the operating system threads used in scaling tests to 
support the execution of HPX threads was limited to 28 (vs. 32 cores 
available in testbed) to allow sufficient hardware resources for OS 
activity and certain runtime system functions (such as asynchronous I/O) 
that also require OS threads.

\begin{figure}[htp]
\centering
\includegraphics[width=.95\linewidth]{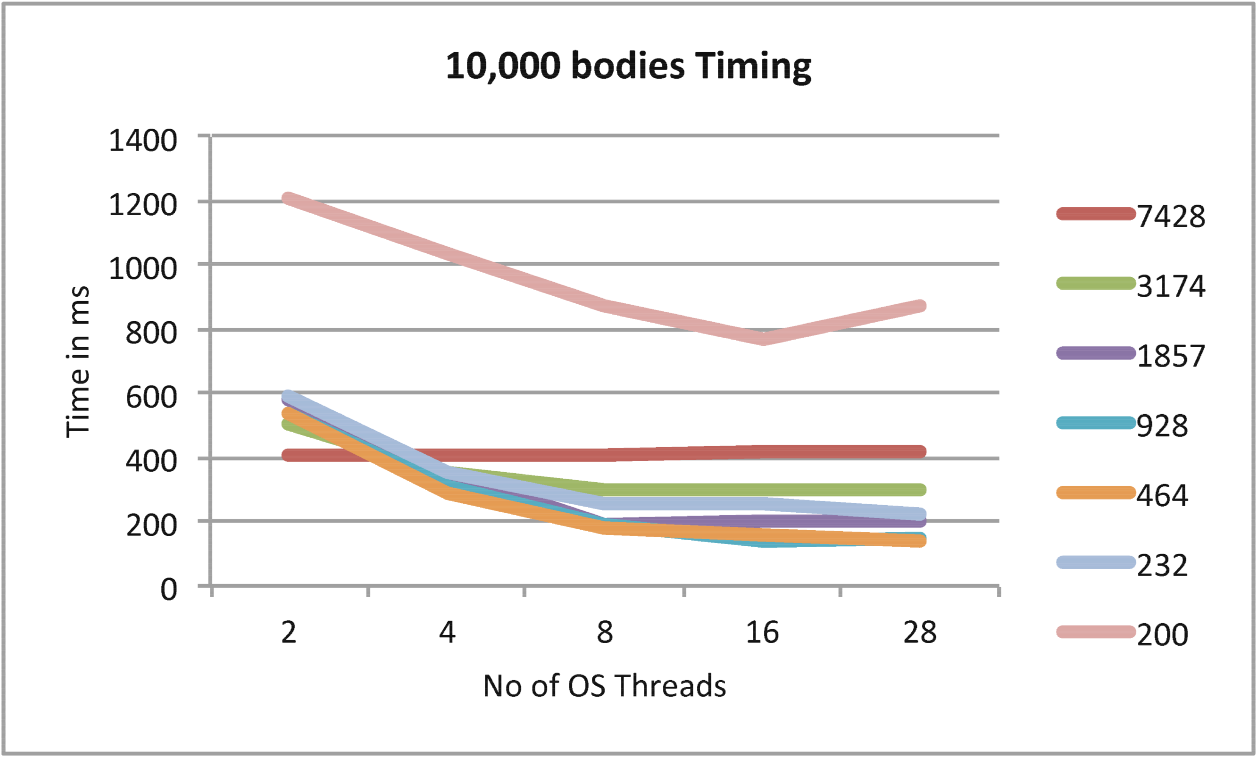}
\caption{Timing of one iteration in the 10,000 particle problem using HPX version of Barnes-Hut N-Body problem
for different grain sizes (indicated by color) over different number of Operating System (OS) threads. As work is made
more fine-grained by decreasing grain size, the number of HPX threads increases. 
While the more fine-grained cases have improved load balancing, they also have more overhead resulting from the
increased number of HPX threads. The optimal grain size for a problem can change as the number of OS threads
increases.}
\label{10000hpxscaling1}\end{figure}

\begin{figure}[htp]
\centering
\includegraphics[width=.95\linewidth]{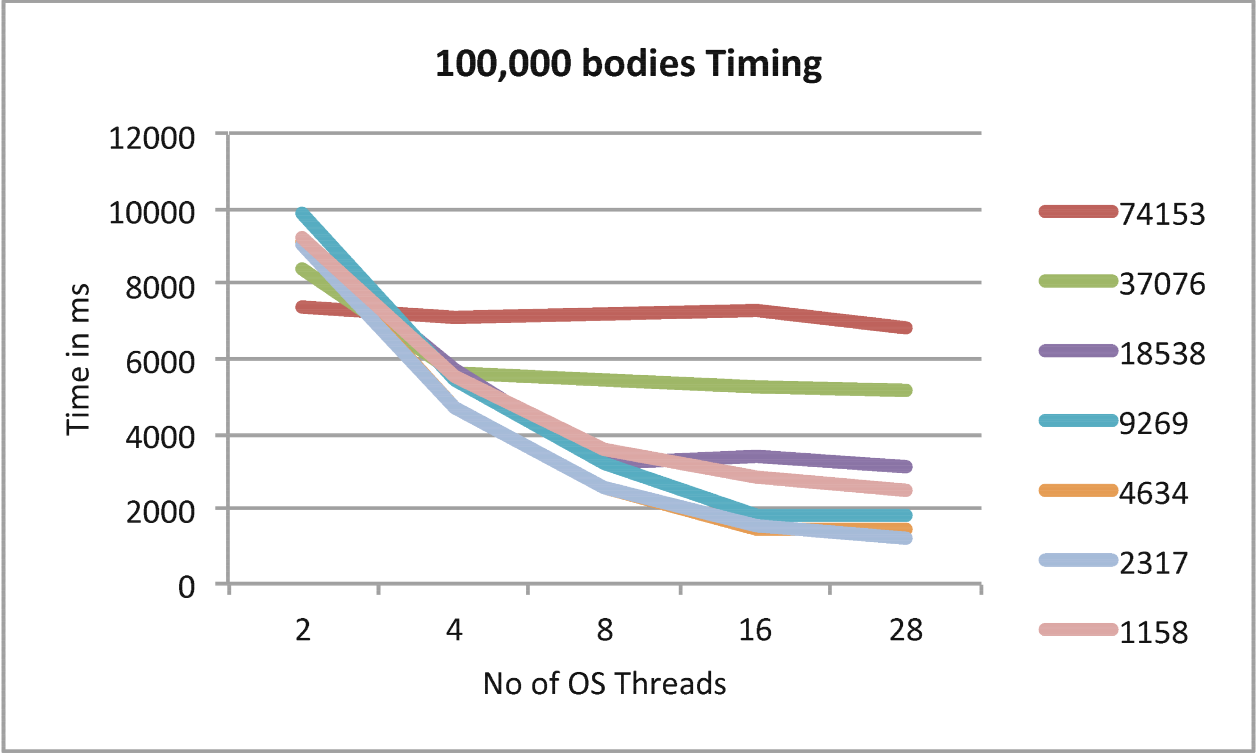}
\caption{Timing of one iteration in the 100,000 particle problem using HPX version of Barnes-Hut N-Body problem
for different grain sizes (indicated by color) over different number of OS threads.
As in Figure~\ref{10000hpxscaling1}, an optimal grain size for a particular number of OS threads exists which
balances the need for fine-grained parallelism and minimal HPX thread overhead.}
\label{100000hpxscaling1}
\end{figure}

\begin{figure}[htp]
\centering
\includegraphics[width=.95\linewidth]{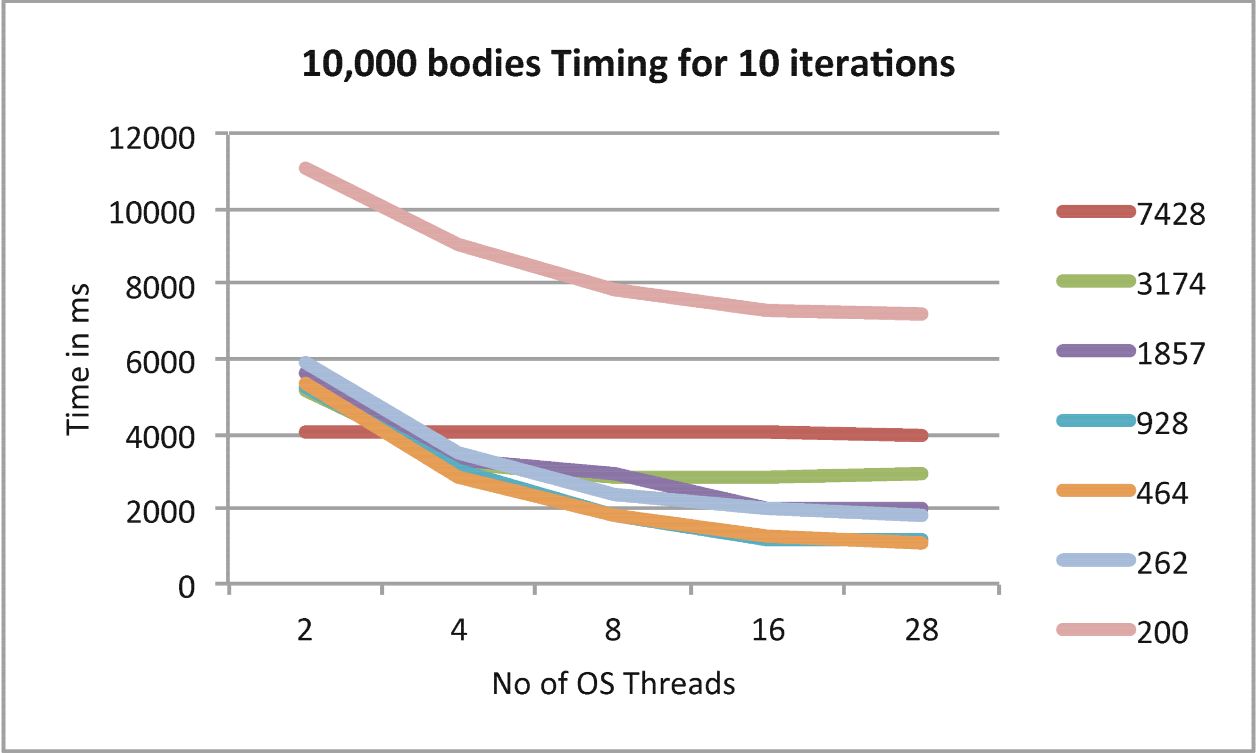}
\caption{Timing of ten iterations in the 10,000 particle problem using HPX version of Barnes-Hut N-Body problem
for different grain sizes (indicated by color) over different number of Operating System (OS) threads.
The same trends seen in Figures~\ref{10000hpxscaling1} and~\ref{100000hpxscaling1} persist as the 
number of iterations of the
N-Body problem increases. An optimal grain size continues to appear which both load balances the problem and 
minimizes thread overhead.}
\label{10000hpxscaling10}
\end{figure}

In Figures~\ref{10000hpxscaling1},~\ref{100000hpxscaling1}, and~\ref{10000hpxscaling10} the trade-off between fine-grained 
parallelism and overhead is explored for cases with various numbers of particles and iterations. In each
example, as the grain size is decreased, the Barnes-Hut N-Body problem is better load balanced 
at a cost of the higher overhead
which results from the addition HPX threads required. The optimal grain size for each problem is found experimentally
and generally differs for different problem sizes and, occasionally, different numbers of OS threads. 

\begin{figure}[htp]
\centering
\includegraphics[width=.95\linewidth]{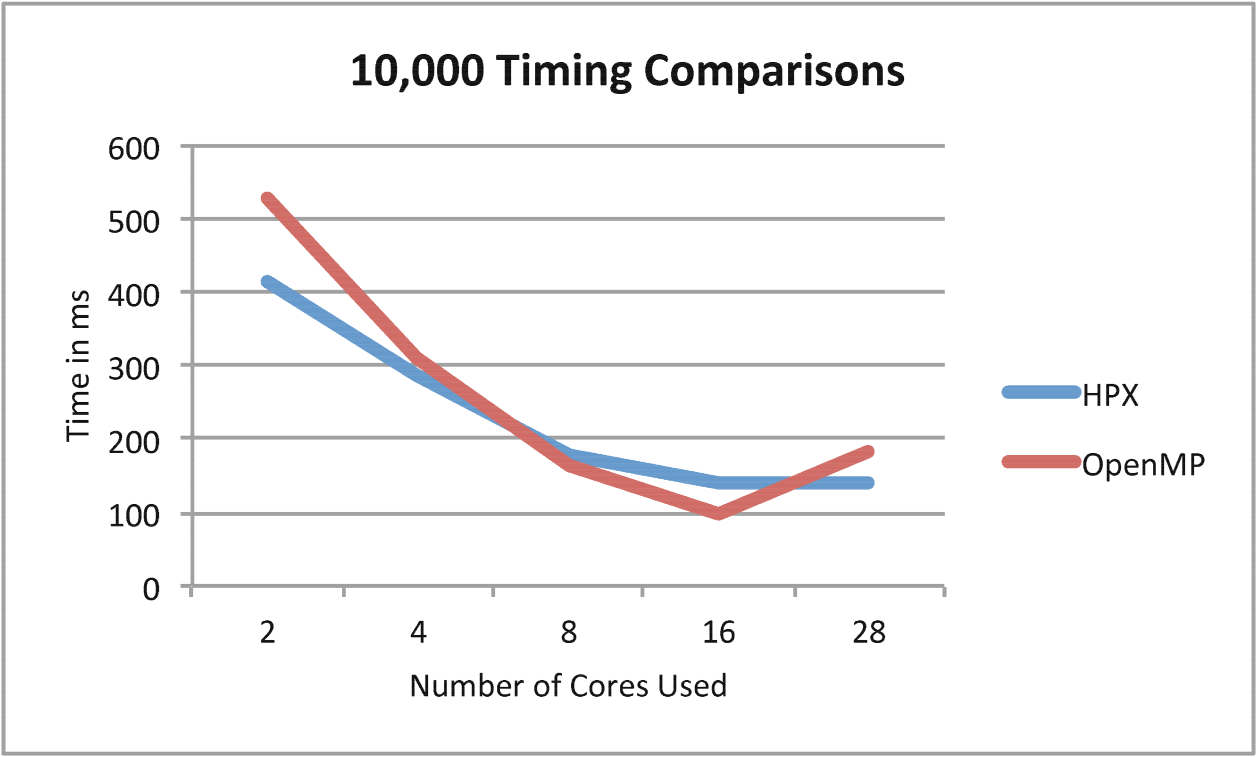}
\caption{Comparison of timing for 10,000 particles in HPX and OpenMP versions of Barnes-Hut N-Body problem for 
multiple iterations where the optimal grain size has been selected for the HPX version of Barnes-Hut. 
For scaling comparison, see Figure~\ref{10komphpx-scaling}.}
\label{10komphpx-timing}
\end{figure}

\begin{figure}[htp]
\centering
\includegraphics[width=.95\linewidth]{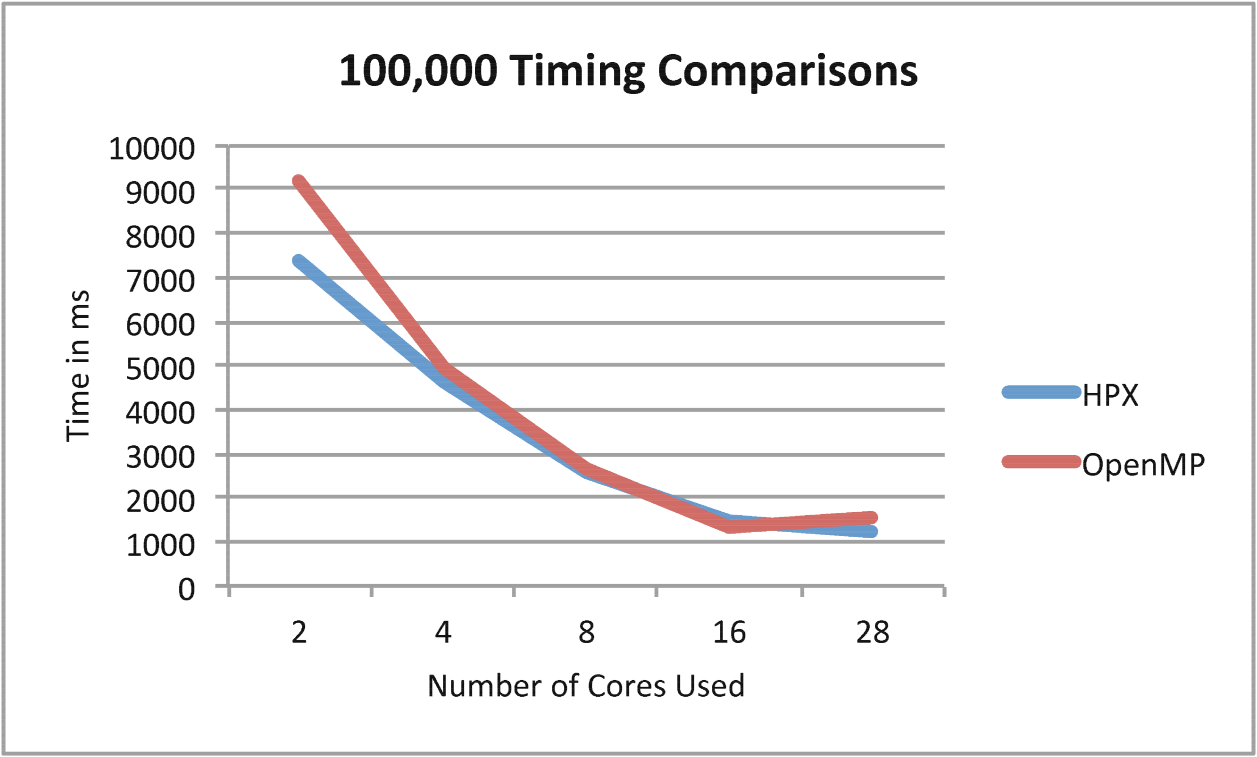}
\caption{Comparison of timing for 100,000 particles in HPX and OpenMP versions of Barnes-Hut N-Body problem for 
multiple iterations where the optimal grain size has been selected for the HPX version of Barnes-Hut. 
For scaling comparison, see Figure~\ref{100komphpx-scaling}.}
\label{100komphpx-timing}
\end{figure}

\begin{figure}[htp]
\centering
\includegraphics[width=0.95\linewidth]{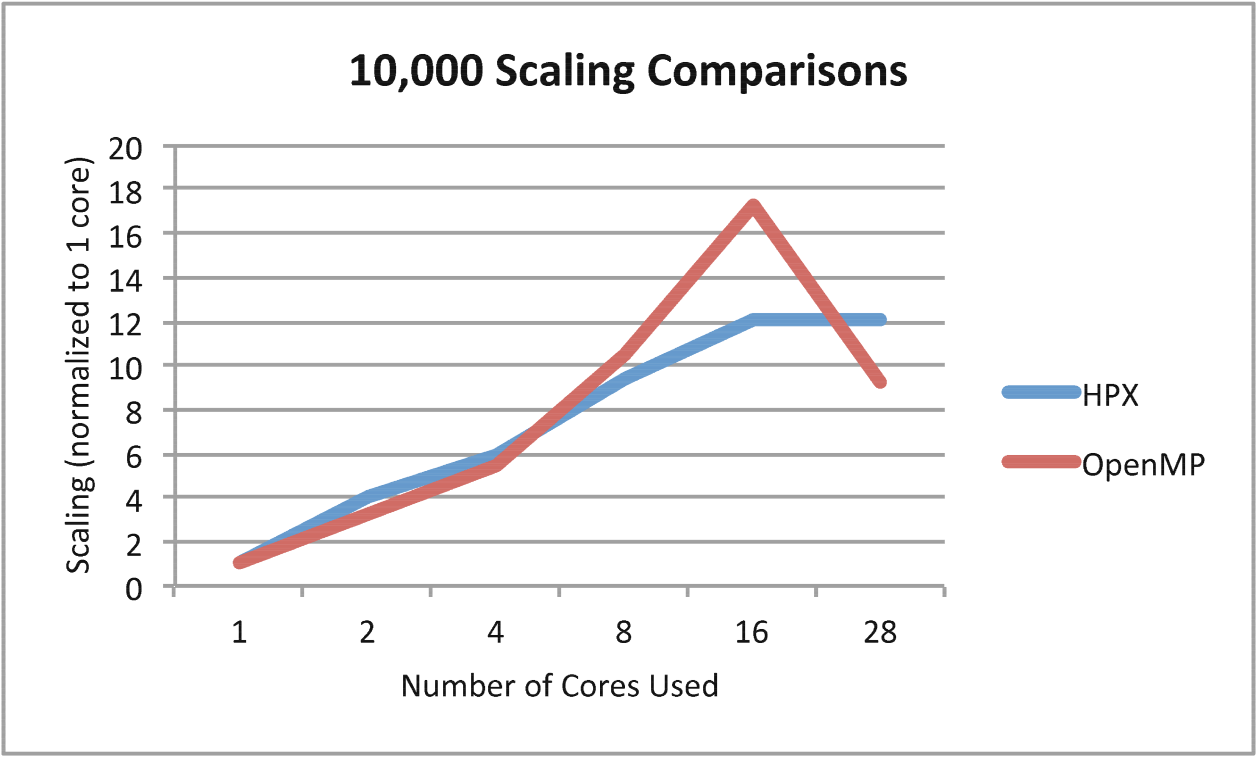}
\caption{Comparison of scaling for 10,000 particles using HPX and OpenMP versions of Barnes-Hut N-Body problem for 
multiple iterations where the optimal grain size has been selected for the HPX version of Barnes-Hut. 
For timing comparison, see Figure~\ref{10komphpx-timing}.}
\label{10komphpx-scaling}
\end{figure}

\begin{figure}[htp]
\centering
\includegraphics[width=.95\linewidth]{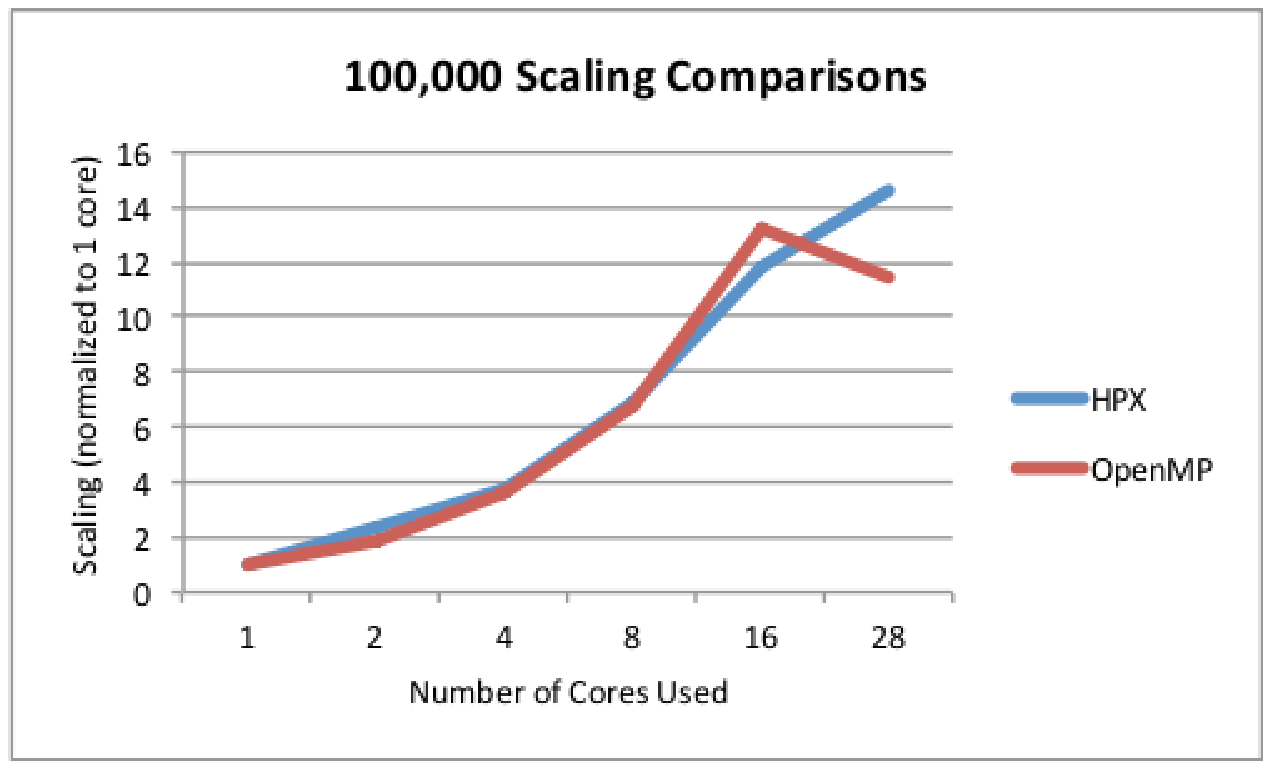}
\caption{Comparison of scaling for 100,000 particles using HPX and OpenMP versions of Barnes-Hut N-Body problem for 
multiple iterations where the optimal grain size has been selected for the HPX version of Barnes-Hut. 
For timing comparison, see Figure~\ref{100komphpx-timing}.}
\label{100komphpx-scaling}
\end{figure}

In Figures~\ref{10komphpx-timing} and~\ref{10komphpx-scaling} the timing comparison 
and scaling comparison between the HPX and
OpenMP versions of Barnes-Hut using multiple iterations are presented in the case of 10,000 particles. 
Figures~\ref{100komphpx-timing} and~\ref{100komphpx-scaling} show similar information for the case of 100,000 particles.
In the case of 100,000 particles, HPX both outscales and outperforms the OpenMP implementation. This is attributed to the improved
load balancing resulting from the dataflow semantics and fine-grained parallelism provided by ParalleX. Also key 
to this result is the flexibility in ParalleX to control the thread grain size as evidenced in  
Figures~\ref{10000hpxscaling1},~\ref{100000hpxscaling1},and~\ref{10000hpxscaling10}. The main caveat for
fine grained load balancing is that the potential for overheads to dominate the computations is higher.

The results obtained from the ParalleX version of the Barnes-Hut algorithm are consistent irrespective of
the number of iterations used or the problem size. The performance, scaling, and the consistent results validate
the viability of ParalleX semantics for variable data structure problems such as the Barnes-Hut N-Body problem.

\section{Conclusion}
\label{conclusion}
The HPX Barnes-Hut implementation presented here is the first implementation 
of a tree based scientific computation using the ParalleX model of execution.
With the overall goal
of providing extreme scalability to scaling-constrained applications, especially to those 
involving trees and graphs, this
work provides a demonstration of techniques to exploit parallelism within each iteration of the 
Barnes-Hut algorithm. 
Direct comparisons were made with an OpenMP implementation of Barnes-Hut in order to provide a benchmark
against which to compare the results. However, it is noted that HPX is not limited to shared memory machines.

The HPX Barnes-Hut implementation exploits futures based asynchrony and AGAS-based
globally unique identifiers to facilitate on-demand data driven force computation. Conventional parallelism approaches
rely on access to data through either system architectures such as shared memory where the data resident in shared
memory can be accessed by all computing resources, or through message passing where, in many cases, the critical
path of execution is blocked until remote data can be fetched.
Future based data access in the HPX Barnes-Hut implementation
allows for local force computations to continue executing operations on
available data while some of the threads are suspended in the {\tt get()}
call.
In the case of remote fetches which cannot be satisfied immediately, the executing HPX thread suspends
operation allowing for another dataflow element to continue force calculation. This on-demand data access provides
asynchrony without blocking the critical execution path, thereby allowing the 
application to fully exploit the underlying resources. It is expected that this feature will be especially useful in a distributed memory setting.

Future extensions of this work will include performance benchmarks on distributed memory machines benchmarked against an
OpenMP--MPI hybrid Barnes-Hut implementation for comparison. 
The performance of current Barnes-Hut implementation in HPX can be further 
improved by applying alternative methods for computation of the 
approximation criterion $\theta$ 
and incorporating multipole moments while removing the global 
barrier so that multiple iterations can proceed simultaneously. 

%
%
\noindent{\bf{\em Acknowledgments:}}
We would like to thank Steven Brandt and Dylan Stark for stimulating discussions.
We acknowledge support comes from NSF grants 1048019 and 1029161 
to Louisiana State University.

%
%
\bibliography{./pxBib}
\bibliographystyle{plain}

%
%
\end{document}